\documentclass[conference]{IEEEtran}
\IEEEoverridecommandlockouts

\usepackage{cite}
\usepackage{amsmath,amssymb,amsfonts}
\usepackage{booktabs}
\usepackage{amsmath}
\usepackage{array}
\usepackage{algorithm}
\usepackage{algorithmic}
\usepackage{graphicx}
\usepackage{textcomp}
\usepackage{comment}
\usepackage{xcolor}
\def\BibTeX{{\rm B\kern-.05em{\sc i\kern-.025em b}\kern-.08em
    T\kern-.1667em\lower.7ex\hbox{E}\kern-.125emX}}
\begin{document}

\title{Enhancing Split Learning with Sharded and Blockchain-Enabled SplitFed Approaches}

\author{\IEEEauthorblockN{Amirreza Sokhankhosh}
\IEEEauthorblockA{\textit{University of Manitoba, Canada}\\
sokhanka@myumanitoba.ca}
\and
\IEEEauthorblockN{Khalid Hassan}
\IEEEauthorblockA{\textit{University of Manitoba, Canada}\\
hassank2@myumanitoba.ca}
\and
\IEEEauthorblockN{Sara Rouhani}
\IEEEauthorblockA{\textit{École de Technologie Supérieure, Canada}\\
sara.rouhani@etsmtl.ca}
}

\maketitle

\begin{abstract}
Collaborative and distributed learning techniques, such as Federated Learning (FL) and Split Learning (SL), hold significant promise for leveraging sensitive data in privacy-critical domains. However, FL and SL suffer from key limitations—FL imposes substantial computational demands on clients, while SL leads to prolonged training times. To overcome these challenges, SplitFed Learning (SFL) was introduced as a hybrid approach that combines the strengths of FL and SL. Despite its advantages, SFL inherits scalability, performance, and security issues from SL. In this paper, we propose two novel frameworks: Sharded SplitFed Learning (SSFL) and Blockchain-enabled SplitFed Learning (BSFL). SSFL addresses the scalability and performance constraints of SFL by distributing the workload and communication overhead of the SL server across multiple parallel shards. Building upon SSFL, BSFL replaces the centralized server with a blockchain-based architecture that employs a committee-driven consensus mechanism to enhance fairness and security. BSFL incorporates an evaluation mechanism to exclude poisoned or tampered model updates, thereby mitigating data poisoning and model integrity attacks. Experimental evaluations against baseline SL and SFL approaches show that SSFL improves performance and scalability by 31.2\% and 85.2\%, respectively. Furthermore, BSFL increases resilience to data poisoning attacks by 62.7\% while maintaining superior performance under normal operating conditions. To the best of our knowledge, BSFL is the first blockchain-enabled framework to implement an end-to-end decentralized SplitFed Learning system.

\end{abstract}


\section{Introduction}

Data privacy is a critical concern in training machine learning models, especially as sensitive data becomes increasingly distributed across decentralized devices. This has led to growing interest in collaborative learning approaches \cite{duan2022combined}, where multiple participants contribute to training a global model while retaining their data locally. By sharing model updates instead of raw data, collaborative learning significantly mitigates privacy risks.

The most widely adopted collaborative learning techniques are Federated Learning (FL) \cite{mcmahan2017communication} and Split Learning (SL) \cite{gupta2018distributed, vepakomma2018split}. In FL, clients train a global model in parallel using their local datasets and share model updates, which are then aggregated by a server using methods like FedAvg \cite{ZHANG2021106775}. However, FL imposes substantial computational demands on clients, limiting its feasibility for resource-constrained devices \cite{Batool2022BlockFeST}.
To address this, SL introduces a model-splitting approach, dividing the global model into smaller client-side and larger server-side segments. Clients train the smaller segment with fewer layers, while the server handles the larger segment, thereby reducing computational overhead on clients. Despite this advantage, SL suffers from prolonged training times due to sequential client-server interactions and frequent communication for each batch \cite{duan2022combined, Gao2020endtoend, ceballos2020splitnn}. Moreover, SL achieves lower efficiency compared to FL, as it does not fully exploit aggregation techniques like FedAvg \cite{thapa2022splitfed}.

To tackle these challenges, SplitFed Learning (SFL) \cite{thapa2022splitfed} was introduced as a hybrid approach that combines the strengths of FL and SL. SFL incorporates an additional federated server to aggregate client-side updates using the FedAvg algorithm, significantly reducing the training time compared to SL. However, as the number of clients increases, the computational burden on the SL server becomes substantial. Additionally, the growing number of clients exacerbates communication overhead, placing a strain on the network infrastructure.

The scalability limitations of both SL and SFL extend beyond infrastructure concerns to performance degradation. As highlighted in \cite{oh2022locfedmix, pal2021server, oh2023mix2sfl}, an imbalance in model updates between the SL server and clients leads to a decline in model performance as the client count increases.

SFL architectures also face severe security challenges. For instance, a malicious FL server could compromise the integrity of the global model by selectively including harmful updates \cite{li2021bflc}. Similarly, malicious clients may disrupt the system by submitting corrupted updates. Moreover, the centralized nature of both servers creates single points of failure, posing significant operational risks as the servers holds all contributions to the global model \cite{qu2022blockchain}. 

To overcome these challenges, we introduce the first end-to-end decentralized and scalable blockchain-enabled SplitFed Learning architecture. We begin our design by first proposing Sharded SplitFed Learning (SSFL), a novel architecture designed to enhance the scalability and performance of SFL by distributing the SL server workload across multiple parallel shards. In SSFL, clients are assigned to separate shards, effectively balancing the computational load and improving training efficiency. SSFL also addresses the issue of imbalanced model updates in SFL by employing an additional federated server that aggregates the shard-level models using the FedAvg algorithm \cite{mcmahan2017communication}. This aggregation effectively ``smooths out" each SL server's model updates, mitigating the adverse effects of high local learning rates and server biases that can impede convergence. By leveraging this approach, SSFL not only improves scalability but also enhances the stability and convergence speed of SFL.

Furthermore, we expand our design by integrating the blockchain architecture with SSFL to resolve persistent security vulnerabilities due to the inherent centralization\cite{pasquini2021unleashing}. In this paper, we propose Blockchain-enabled SplitFed Learning (BSFL), the first decentralized SplitFed Learning architecture that eliminates reliance on a central FL server and fortifies security through a committee-based blockchain consensus mechanism \cite{li2021bflc}. In BSFL, tasks traditionally handled by the central FL server—such as model aggregation, update validation, and reward distribution—are autonomously executed using smart contracts on the blockchain. A decentralized committee of nodes evaluates model updates from each SFL shard, mitigating data poisoning attacks and ensuring the integrity of contributions. This blockchain-based framework creates a trustless, tamper-resistant learning environment, leveraging consensus to reinforce security, fairness, and accountability. 
To summarize, the contributions of this paper are as follows:
\begin{enumerate}
    \item Sharded SplitFed Learning (SSFL): We propose SSFL as an enhancement to the existing SplitFed Learning (SFL) model to overcome its scalability limitations. This approach reduces computational strain on the SL server by enabling parallel training from multiple SFL shards, thus significantly improving the scalability and efficiency of the model. SSFL also addresses the issue of imbalanced model updates by reducing the effective learning rate through federated averaging. SSFL is particularly beneficial in environments with a large number of resource-constrained devices.
    \item Blockchain-enabled SplitFed Learning (BSFL): We introduce BSFL, the first blockchain-enabled SFL framework, to address the centralization-related security vulnerabilities in SFL. BSFL replaces the centralized FL server with a blockchain architecture employing a committee-based consensus mechanism. It incorporates an evaluation process to assess client model updates, thereby mitigating risks such as data poisoning and model tampering while enhancing system fairness, robustness, and performance.
    \item Experiments and Evaluation: We demonstrate the effectiveness of SSFL and BSFL through comprehensive experiments designed to measure performance under both standard operating conditions and simulated data-poisoning attacks. Additionally, we compare the round completion time and convergence speed of our approaches with traditional SL and SFL models, demonstrating superior performance and operational efficiency.
\end{enumerate}

The remainder of this paper is organized as follows. In Section \ref{sec:related-works}, we discuss previous studies related to scalable split learning and blockchain-enabled distributed learning. Section \ref{sec:def} introduces definitions used throughout the paper. In Sections \ref{sec:ssfl} and \ref{sec:bsfl}, we introduce SSFL and BSFL, respectively. Section \ref{sec:discussions} provides detailed discussions for our proposed frameworks. Section \ref{sec:experiments} presents our comprehensive evaluation and results. At last, Sections \ref{sec:future-works} and \ref{sec:conclusion} address the future works and conclusion.

\section{Related Works}
\label{sec:related-works}

FL was first proposed by McMahan et al. \cite{mcmahan2017communication} to enhance the privacy of data holders in machine learning training procedures. In FL, a network comprising clients containing local datasets and a server capable of communicating with all clients is established. In this scenario, clients contribute to training a selected global model by sharing model updates instead of their sensitive data to preserve their right to privacy \cite{kairouz2021advances}. In FL, clients train a global model using their local datasets, which requires substantial computational resources. To address this limitation, Gupta and Raskar \cite{gupta2018distributed} introduced SL. In SL, the machine learning model is divided into smaller segments: clients train the smaller segments locally, while the server performs the heavier computational tasks. However, SL involves significant communication overhead between clients and the server, requiring two messages per batch for each client. Consequently, the convergence time of SL is much slower compared to FL.

SFL is introduced by Thapa et al. \cite{thapa2022splitfed} as a new collaborative and distributed learning framework that combines the strengths of FL and SL to overcome the limitations of both approaches. SFL operates as a variation of SL, where clients train their model segments and communicate with the server in parallel. Additionally, client model updates are aggregated using the FedAvg algorithm by a federated server. By integrating FL into the SL architecture, SFL significantly improves the speed and efficiency of the original SL framework.

\subsection{Scalibility}
The inherent model split nature of SL and SFL has been identified as a key factor limiting their scalability \cite{oh2022locfedmix}. In parallel SL, the server model segment is updated significantly more often than the client segments, resulting in an uneven learning rate between the two. This imbalance can lead to inefficient training, especially when client model updates are handled separately from the server model. To address these challenges, LocFedMix-SL was introduced, incorporating local regularization of client models and augmenting clients' smashed data to improve server model training \cite{oh2022locfedmix}.
Building on this work, additional solutions have been proposed. For instance, Pal et al. \cite{pal2021server} addressed the same issues by implementing different learning rates for the server and clients, along with broadcasting similar gradients to all clients. Later, Mix2SFL combined these approaches to achieve optimized performance, ensuring both enhanced accuracy and improved communication efficiency \cite{oh2023mix2sfl}.
Although these methods effectively tackle performance issues related to scalability, they fall short in addressing the computational and communication overhead of the server and its impact on overall efficiency. To bridge this gap, we propose SSFL, a novel sharding approach designed to enhance the accuracy, stability, and scalability of SFL.

\subsection{Security}
The intersection of blockchain technology and FL has emerged as a promising avenue for addressing persistent challenges in secure and privacy-preserving machine learning. FL enhances data privacy by facilitating distributed training and aggregating model updates instead of collecting private datasets. However, despite its privacy-preserving design, security vulnerabilities persist in the FL architecture \cite{lyu2020threats}.
One major concern arises from malicious clients, who may intentionally transmit harmful model updates to degrade the performance of the global model. Similarly, the centralized server in FL introduces risks of unfairness, as it may selectively aggregate specific model updates, potentially biasing the global model. Moreover, the server is susceptible to performing Membership Inference Attacks (MIA) \cite{shokri2017membership} on the received model updates, enabling it to infer details about clients' local datasets.
Although employing differential privacy \cite{dwork2006differential} mitigate the risk of MIAs, this approach often comes at the cost of reduced model performance and efficiency \cite{li2020fl}. As a result, addressing these security challenges without compromising the effectiveness of FL remains an open area of research, prompting the exploration of blockchain as a potential solution. 

To resolve these issues, blockchain-enabled federated learning approaches were suggested \cite{qu2022blockchain}. The decentralized feature of blockchains eliminate the challenges related to the server component in FL. In addition, the inherent incentive mechanisms utilized by blockchains contribute to encouraging local clients to train truthfully and communicate honest model updates to receive greater rewards. Kim et al. \cite{kim2020blockfl} introduced the first blockchain-enabled federated learning framework called BlockFL. Following their proposal, Li et al. \cite{li2021bflc} introduced a blockchain-based FL blueprint utilizing a committee consensus. The committee consensus used in our design is motivated by this study. A similar committee consensus has also been incorporated in Proof-of-Collaborative-Learning \cite{Sokhankhosh2024PoCL}, which leverages blockchain computation power to distributedly train FL models. The integration of blockchain with FL extends beyond committee consensus mechanisms and includes various enhancements, such as its combination with differential privacy \cite{truex2019hybrid, qu2021proof, Lu2020PoTQ}, Layer 2 (L2) solutions \cite{Yuan2021ChainsFL, Alief2023FLB2}, and Zero-Knowledge Proofs (ZKPs) \cite{zhang2023blockchain}. These approaches collectively aim to enhance privacy, scalability, and security in blockchain-enabled FL frameworks.

The integration of blockchain networks with SL remains relatively underexplored. BlockFeST, proposed by Batool et al. \cite{Batool2022BlockFeST}, combines blockchain with federated and split learning, offloading major computational responsibilities to an SL server to alleviate the computational burden on FL clients. Similarly, Sai \cite{Sai2024Ablockchain} introduced a blockchain-enabled split learning framework that leverages smart contracts to dynamically select servers and clients for each training round. 
\begin{figure}[h]
  \centering
  \includegraphics[width=\linewidth]{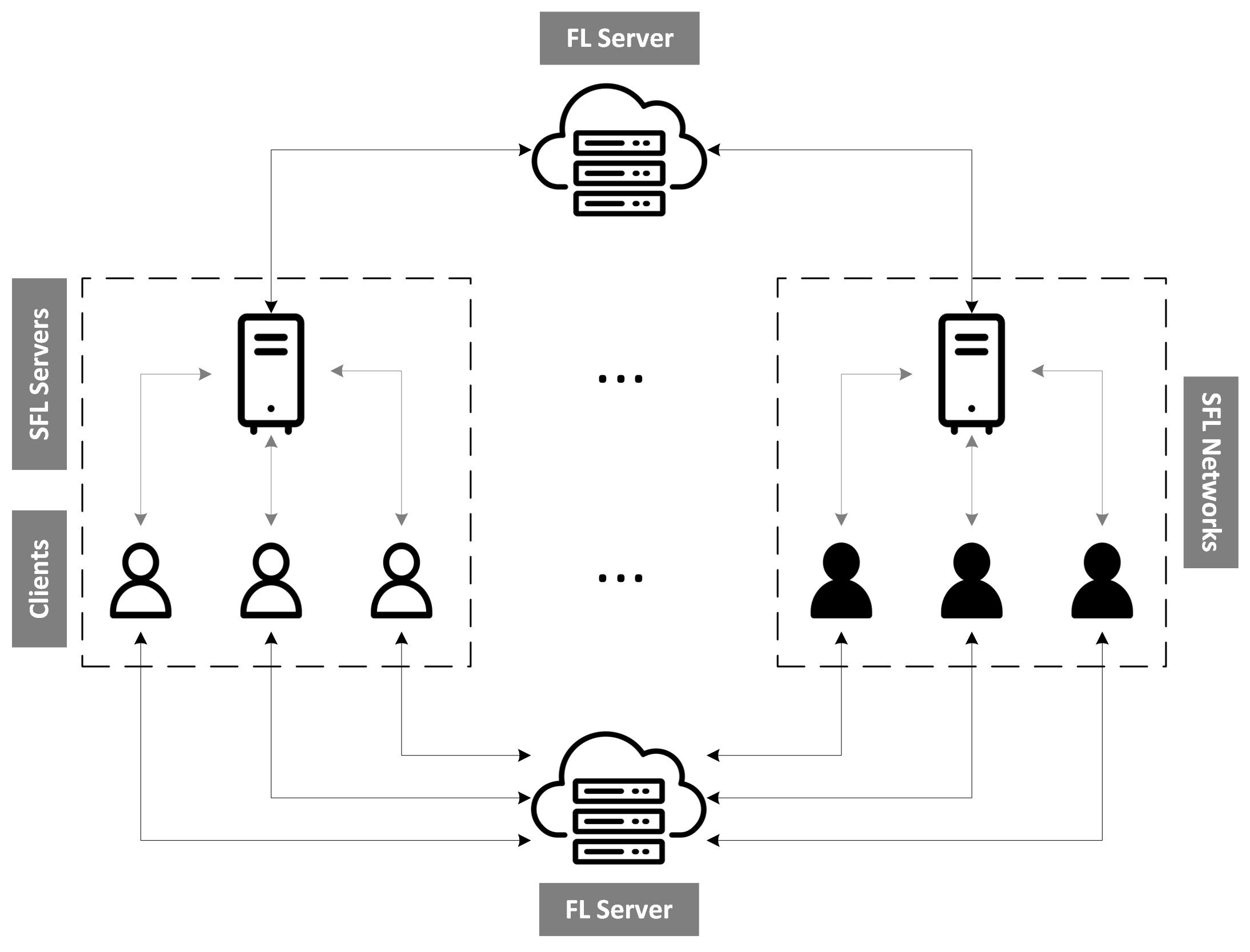}
  \caption{Overview of Sharding SplitFed Learning framework.}
  \label{fig:ssfl}
\end{figure}
While both approaches improve fairness and security, they continue to face significant challenges related to centralization and scalability.
Hence, to the best of our knowledge, we are the first to propose Blockchain-enabled SplitFed Learning to eradicate the security, fairness, and performance concerns of Split Learning. 
\section{Definitions}
\label{sec:def}
To enhance the readability of the paper, let us define a few incorporated terms of our design.

\textbf{Client}: A client is any node in a decentralized learning algorithm that possesses private data and contributes to the training process by training a model locally on its dataset.

\textbf{Server}: A server is any node that participates in distributed training without holding private data. In Split Learning (SL), the server trains the computationally intensive portion of the model. In Federated Learning (FL), the server solely aggregates client models to construct a new global model. Since the SFL, SSFL, and BSFL algorithms include both Split Learning and Federated Learning servers, their distinction is vital for understanding the algorithms presented in the paper.

\textbf{Node}: A node refers to any participant in the training process, either a client or a server (SL or FL).

\section{Sharded SplitFed Learning}
\label{sec:ssfl}

As a variation of SL, SFL combines the beneficial characteristics of both SL and FL to address their respective limitations. In the SFL architecture, two primary innovations are introduced: parallel client training and a federated server that aggregates client models at each training round. While these modifications significantly reduce SL’s convergence time, scalability remains a major challenge. In both SFL and SL, as the number of clients increases, the computational and communication overhead on the central SL server grows substantially, limiting the feasibility of these architectures in large-scale, real-world applications.

\begin{algorithm} 
\caption{Sharded SplitFed Learning (SSFL)} 
\begin{algorithmic}[1] 
\label{alg:ssfl} 
\STATE \textit{/* Executes on each Shard Server (SFL Server) */} 
\ENSURE \textbf{TrainingCycle}($c$): 
\FOR{each round $r = 1, 2, \dots, R$} 
    \FOR{each client $j \in \text{shard}$ in parallel} 
        \FOR{each batch $b$ in epoch $E$} 
            \STATE $(A_{j,r}, Y_{j,r}) \leftarrow \text{ClientTrain}(W^C_{j,r})$ 
            \STATE $\hat{Y}_{j,r} \leftarrow \text{ServerForwardPass}(A_{j,r}, W^S_{i,j,r})$
            \STATE $\mathcal{L}_{j,r} \leftarrow \text{Loss}(Y_{j,r}, \hat{Y}_{j,r})$ 
            \STATE $\nabla \ell_j(W^S_{i,j,r}, A_{j,r}) \leftarrow \text{ComputeGradients}(\mathcal{L}_{j,r})$ 
            \STATE $W^S_{i,j,r} \leftarrow W^S_{i,j,r} - \lambda \cdot \nabla \ell_j(W^S_{i,j,r})$ 
            \STATE Send $d A_{j,r} := \nabla \ell_j(W^S_{i,j,r}, A_{j,r})$ to client $j$. 
        \ENDFOR 
    \ENDFOR 
    \STATE Update shard server model:
    \STATE $W^S_{i,r+1} \leftarrow \frac{1}{J} \sum_{j=1}^{J} W^S_{i,j,r}$ 
\ENDFOR
\STATE{}
\STATE \textit{/* Executes on the Federated Learning (FL) Server */}
\ENSURE \textbf{Training}():
    \FOR{each training cycle $t = 1, 2, \dots, T$}
        \FOR{each shard server $i \in \text{SFL network}$ in parallel}
            \STATE Call \textbf{TrainingCycle}($t$) for shard $i$.
        \ENDFOR
    \ENDFOR
\STATE{}
\STATE \textit{/* Executes on the Federated Learning (FL) Server */}
\ENSURE \textbf{Aggregate}():
\STATE Receive $W^S_{i,t}$ and $W^C_{j,t}$ updates from each shard server $i$ and client $j$.
\STATE Update global models:
\STATE $W^S_{t+1} \leftarrow \frac{1}{I} \sum_{i=1}^{I} W^S_{i,t}$ 
\STATE $W^C_{t+1} \leftarrow \frac{1}{J} \sum_{j=1}^{J} W^C_{j,t}$ 
\end{algorithmic}
\end{algorithm}

However, the scalability challenges in SL and SFL are not solely limited by infrastructure capacity. SL and its variations experience a performance degradation that intensifies as the number of clients increases. In these frameworks, the server-side model is updated far more frequently than the client-side model on each local device, resulting in an imbalance in model training. Consequently, as the client count grows, this disparity between server and client models leads to a substantial decline in overall performance. While previous studies have proposed various solutions to address the performance and efficiency limitations of SL and SFL in scalable environments \cite{oh2022locfedmix, pal2021server, oh2023mix2sfl}, to the best of our knowledge, no work has comprehensively addressed the imbalanced training problem, as well as the challenges of server-side computation and communication overhead.

\begin{table}[h!]
\centering
\caption{Notations}
\label{table:notation}
\resizebox{\columnwidth}{!}{%
\begin{tabular}{@{}p{0.25\columnwidth} p{0.65\columnwidth}@{}}
\toprule
\textbf{Symbol} & \textbf{Description} \\ 
\midrule
\( R \) & Total number of training rounds in a shard server \\
\( T \) & Total number of training cycles \\
\( E \) & Number of epochs for each client’s local training \\
\( J \) & Number of clients per shard \\
\( I \) & Total number of shard servers in the SSFL network \\
\( K \) & Number of top committee-selected model updates in BSFL \\
\( W^S_{i,j,r} \) & Model weights of shard server \( i \) for client \( j \) at round \( r \) \\
\( W^C_{j,r} \) & Local model weights of client \( j \) at round \( r \) \\[0.15cm]
\( W^S_{t} \) & Server global model at cycle \( t \) \\[0.1cm]
\( W^C_{t} \) & Client global model at cycle \( t \) \\
\( X_j \) & Input data for client \( j \) \\
\( Y_j \) & True labels associated with \( X_j \) \\
\( A_{j,r} \) & Activation output from client \( j \) at round \( r \) \\
\( \hat{Y}_{j,r} \) & Predicted labels for client \( j \) at round \( r \) \\
\( \mathcal{L}_{j,r} \) & Loss computed for client \( j \) at round \( r \) \\
\( \nabla \ell_j(W^S_{i,j,r}, A_{j,r}) \) & Gradient of the loss function with respect to \( W^S_{i,j,r} \) and \( A_{j,r} \) \\
\( dA_{j,r} \) & Feedback gradient (from shard server to client) for client \( j \) at round \( r \) \\
\( \nabla \ell_j(W^C_{j,r}) \) & Gradient of the loss function with respect to the local model weights \( W^C_{j,r} \) for client \( j \) at round \( r \) \\
\( \lambda \) & Learning rate for updating local client models \\
\bottomrule
\end{tabular}%
}
\end{table}

To tackle these challenges, we propose Sharded SplitFed Learning (SSFL), an enhanced framework where clients are organized across multiple shard servers, each acting as a conventional SplitFed Learning SL server. The SSFL architecture, illustrated in Figure \ref{fig:ssfl}, introduces an additional federated server that aggregates model updates from all shard servers in parallel. Although we depict the two FL servers as separate entities, their responsibilities can be handled by a single instance if the framework implements specific security measures, such as encrypted model updates for clients and servers. For simplicity in our explanation, we assume that these precautions are taken, and the system is only comprised of one FL server aggregating both client and SL server models of the SplitFed Learning networks.

\subsection{SSFL Workflow}
To start training, clients join an SFL network within the system by sending a handshake request to one of the available SFL servers. Upon joining, clients begin training by performing a forward pass on the global client model using their local data. Each batch's output and target are sent to the designated SFL server to continue training. Servers then conduct a forward pass on the received outputs, progressing through the split layer until reaching the final layer, where they compute gradients and update the local SFL server model. The gradient of the final layer is also sent back to the clients, enabling clients to update their respective models. Once all shards complete their local training and aggregation, both clients and SFL servers send their updated models to the FL server for a final aggregation step. Each FL aggregation completes a \textbf{cycle} of training in SSFL. To formalize the SSFL training process, we present Algorithms \ref{alg:ssfl} and \ref{alg:clients}, which detail each workflow component in sequence. Table \ref{table:notation} is an accompanying notation table that provides clear definitions of the variables, parameters, and functions utilized within these algorithms, serving as a reference to streamline understanding.

\subsection{Scalability and Performance}

\begin{algorithm} 
\caption{Client Training in SplitFed Learning} 
\begin{algorithmic}[1] 
\label{alg:clients} 
\STATE \textit{/* Executes on each client device */} 
\ENSURE \textbf{ClientTrain}(): 
\FOR {each batch $b$ in epoch $E$} 
    \STATE $A_{j,r} \leftarrow \text{ClientForwardPass}(X_j, W^C_{j,r})$ 
    \STATE Assuming $Y_j$ is the label of $X_j$
    \STATE Send $(A_{j,r}, Y_{j})$ to the assigned shard server 
\ENDFOR
\STATE{}
\STATE \textit{/* Executes on each client after receiving the gradients */}
\ENSURE \textbf{ClientBackProp}():
\STATE Receive $d A_{j,r}$ from shard server.
\STATE $\nabla \ell_j(W^C_{j,r}) \leftarrow \text{BackPropagate}(d A_{j,r})$ 
\STATE $W^C_{j,r} \leftarrow W^C_{j,r} - \lambda \cdot \nabla \ell_j(W^C_{j,r})$ 
\end{algorithmic}
\end{algorithm}

Incorporating shards into the SFL network significantly reduces the computational load on individual SFL servers by distributing the forward and backward propagation tasks across multiple servers. This sharding approach not only enhances scalability by effectively expanding the system’s infrastructure capacity but also improves performance under increasing client numbers. In addition, the sharding mechanism addresses the performance-level scalability challenge, which arises due to the imbalance in learning rates between the SFL server and client models.

As discussed by \cite{oh2022locfedmix}, this performance issue stems from the SFL server's effective learning rate being higher than that of the clients, leading to disproportionately faster updates in the SFL server model compared to client models. This discrepancy results in unstable and suboptimal training as the number of clients increases. To address this imbalance, studies like LocFedMix \cite{oh2022locfedmix} and SGLR \cite{pal2021server} propose solutions such as augmenting smashed client data or adjusting the learning rate to align client and server updates better.

Instead, in our approach, we propose introducing an additional federated server to aggregate updates from shard SFL servers. By using this extra FL server layer, we reduce the learning rate of the shard servers, creating a more balanced and synchronized training process. This adjustment not only mitigates the risk of performance degradation but also ensures that both SFL server and client models progress at a harmonious rate, thereby enhancing the overall stability and efficiency of the training framework.

\begin{algorithm} 
\caption{Blockchain-enabled SplitFed Learning (BSFL)} 
\label{alg:bsfl} 
\begin{algorithmic}[1] 
    \STATE \textit{/* Runs on each SFL server /} 
    \ENSURE \textbf{TrainingCycle}($c$): 
    \FOR{each round $r = 1, 2, \dots, R$} 
        \FOR{each client $j \in \text{shard}$ in parallel} 
            \FOR{each batch $b$ in epoch $E$} 
                \STATE $(A_{j,r}, Y_{j,r}) \leftarrow \text{ClientTrain}(W^C_{j,r})$ 
                \STATE $\hat{Y}_{j,r} \leftarrow \text{ServerForwardPass}(A_{j,r}, W^S_{i,j,r})$
                \STATE $\mathcal{L}_{j,r} \leftarrow \text{Loss}(Y_{j,r}, \hat{Y}_{j,r})$ 
                \STATE $\nabla \ell_j(W^S_{i,j,r}, A_{j,r}) \leftarrow \text{ComputeGradients}(\mathcal{L}_{j,r})$
                \STATE Send $d A_{j,r} := \nabla \ell_j(W^S_{i,j,r}, A_{j,r})$ to each client $j$.    
            \ENDFOR 
        \ENDFOR 
        \STATE Update shard server model:
        \STATE $W^S_{i,r+1} \leftarrow \frac{1}{J} \sum_{j=1}^{J} W^S_{i,j,r}$ 
    \ENDFOR
    \STATE Send $W^S_{i,R}$ to blockchain ledger. 
    \STATE Request each client $j \in \text{shard}$ to submit $W^C_{j,R}$ to ledger.     
    \STATE{} 
    \STATE \textit{/* Runs on each SFL server */} 
    \ENSURE \textbf{Evaluate}($W^S_{i,R}$, [$W^C_{j,R}$ for each client $j$]):
    \STATE Initialize \textbf{scores} as an empty list.
    \FOR{each client $j \in \text{shard}$ in parallel}
        \STATE $A_{j,R} \leftarrow \text{ClientForwardPass}(X, W^C_{j,R})$
        \STATE $\hat{Y}_{j,R} \leftarrow \text{ServerForwardPass}(A_{j,R}, W^S_{i,R})$ 
        \STATE $\mathcal{L}_{j,R} \leftarrow \text{Loss}(Y_{j,R}, \hat{Y}_{j,R})$
        \STATE Append $\mathcal{L}_{j,R}$ (validation loss) to \textbf{scores}.
    \ENDFOR
    \RETURN \textbf{Median}(\textbf{scores})    
    \STATE{}
    \STATE \textit{/* Runs on the blockchain network */} 
    \ENSURE \textbf{Committee Procedure}: 
    \FOR{each cycle $t = 1, 2, \dots, T$} 
        \IF{$t = 1$}
            \STATE Randomly select committee members.
        \ELSE    
            \STATE Select committee members based on scores from the previous cycle.
        \ENDIF
        \FOR{each committee member $i$ in parallel}
            \STATE \textbf{TrainingCycle}($t$)
        \ENDFOR
        \STATE Receive all $W^S_{i,R}$ and $[W^C_{j,R}]$ updates.
        \FOR{each member $i$}
            \STATE Send $W^S_{i,R}$ and $[W^C_{j,R}]$ to other committee members.
            \STATE Call \textbf{Evaluate} to obtain validation scores.
        \ENDFOR
        \STATE Assign the median of all scores given to shard $i$ as the final score.
        \STATE Sort scores and select top $k$ model updates.
        \STATE Update global models:
        \STATE $W^S_{t+1} \leftarrow \frac{1}{K} \sum_{k=1}^{K} W^S_{k,t}$
        \STATE $W^C_{t+1} \leftarrow \frac{1}{K \cdot J} \sum_{k=1}^{K} \sum_{j=1}^{J} W^C_{k,j,t}$ 
    \ENDFOR
\end{algorithmic} 
\end{algorithm}

\section{Blockchain-enabled SplitFed Learning}
\label{sec:bsfl}

While sharding enhances both the performance and scalability of the SSFL framework, centralization at the server side still introduces several security and reliability risks: (i) The FL server may exhibit bias by favouring certain clients, either intentionally or due to external manipulation, which can alter the training loop, undermining both model integrity and overall training efficiency. (ii) The central FL server in SSFL and its parent systems handles all core functionalities and bears the main workload, making it a critical single point of failure. Any FL server failure would not only halt the system’s operation but also risk the loss or corruption of all contributions, severely impacting system reliability and continuity. (iii) Malicious clients within the SSFL framework may engage in data poisoning attacks, submitting manipulated data to compromise the quality, reliability, and generalizability of the global model. (iv) Centralization increases the risk of privacy breaches, as the FL server handles sensitive information from all clients. 
A breach or malicious access to the central FL server could expose client data, violating privacy and confidentiality requirements. (v) The framework assumes the central FL server is trustworthy and operates without malice. However, if the server is compromised or behaves maliciously, such as through collusion with external entities, it could jeopardize the system’s security, fairness, and trustworthiness.

To address these issues, we propose decentralizing the SSFL framework through the blockchain technology to eliminate the centralized FL server and its associated security risks. In Blockchain-enabled SplitFed Learning (BSFL), smart contracts on the blockchain assume the central server's functionalities, enabling end-to-end decentralization. This paper introduces a committee consensus mechanism within the BSFL framework, responsible for block creation as well as the evaluation and validation of model updates, to ensure secure, unbiased, and robust operations. By designing SSFL and decentralizing it using BSFL, we introduce the first Blockchain-enabled SplitFed Learning system, enhancing fairness, efficiency, and scalability of the original SFL algorithm.

To decentralize the tasks of FL servers, we initialize the global client and SFL server models directly on the blockchain. This setup allows SFL servers and clients to access the global models and commence training. After certain SplitFed training rounds, each client and SFL server saves their updated models on the blockchain, which subsequently triggers an aggregate smart contract to combine these updates into a new global model.

\subsection{Committee Consensus Mechanism}
Consensus mechanisms in blockchains determine the content of blocks and their order within the chain. In this paper, we employ a committee consensus mechanism to evaluate model updates proposed by each shard, selecting only the top-performing models for aggregation. In BSFL, the SFL servers in each shard form a committee that validates each other's model updates by assigning a score to each update during each training cycle. By limiting communication to the committee nodes, the system significantly reduces communication overhead. The scores assigned to committee members are then ranked, and only the top $K$ models are aggregated to create the updated global models. At the start of each new cycle, a new committee is selected based on the scores from the previous cycle, ensuring that prior committee members do not serve consecutively, therefore reducing the risk of collusion or malicious actions.

Each committee member validates the model updates proposed by others by evaluating their data against the newly trained models. The validation loss, or accuracy if preferred, is reported as the score assigned by one member to another. The final score for each committee member in that cycle is determined by selecting the median of all scores received.

The evaluation metric can be based on either validation loss or validation accuracy, depending on the task. The primary distinction lies in the optimization goal: minimizing validation loss versus maximizing validation accuracy. Additionally, it is important to note that validation accuracy can only be utilized in classification problems, whereas validation loss can be employed across a broader range of tasks, including non-classification problems. 

\subsection{BSFL Workflow}
BSFL supervises the training procedure by utilizing three distinct smart contracts to perform the tasks of the central FL server. In the initial round, the \emph{AssignNodes} smart contract randomly selects nodes in the system to represent the SFL servers within each shard. Subsequently, each server is randomly assigned a set of clients to establish the composition of each shard. The training procedure in each SFL network follows the traditional SFL algorithm; however, upon completion, the models are stored on the blockchain. The \emph{ModelPropose} smart contract collects and distributes all trained models to each committee member, i.e., each SFL server. The SFL servers then validate the newly trained global models using their local data and report the validation loss as the score for the models of other members. The median of all scores assigned to a member's model is selected as the final score. The \emph{EvaluationPropose} smart contract sorts these scores and identifies the top $K$ models as the winners of the current cycle, with the aggregate of these models forming the new global models for the subsequent cycle. Algorithm \ref{alg:bsfl} and \ref{alg:clients} illustrate the BSFL workflow. 

\subsection{Node Assignment}
The committee for each new cycle is selected based on the scores of nodes from the previous round. To ensure fairness and enhance security, any committee member from round $t$ is not allowed to remain on the committee in round $t+1$. Thus, we designate a node as an SFL server if its score surpasses those of other nodes not already chosen as servers in the previous round. Shard nodes are assigned sequentially: an SFL server and its clients are first allocated for one shard, followed by the next. Since committee members from the previous cycle must take on client roles in the current cycle, we only consider these previous committee members when assigning clients.

The algorithm for assigning nodes as clients and SFL servers may vary from our proposed method. Our approach groups nodes with similar efficiency within the same shard, ensuring that even in the case of data-poisoning attacks, the performance of honest nodes is less likely to be overshadowed

\begin{table}[h!]
\centering
\caption{Client and Server Model Architectures. \\ \((D \times H \times W)\) are the dimensions of input data.}
\resizebox{\columnwidth}{!}{
\label{table:model}
\begin{tabular}{@{}p{0.2\columnwidth} p{0.2\columnwidth} p{0.25\columnwidth} p{0.25\columnwidth}@{}}
\toprule
\textbf{Model} & \textbf{Layer Type} & \textbf{Parameters} & \textbf{Output Shape} \\ 
\midrule
\textbf{Client Model} & Conv2d & In: $D$, Out: 32, Kernel: \(3 \times 3\) & \(32 \times H \times W\) \\
                           & ReLU   & - & \(32 \times 28 \times 28\) \\
                           & MaxPool2d & Kernel: \(2 \times 2\), Stride: 2 & \(16 \times H/2 \times W/2\) \\
\midrule
\textbf{Server Model} & Conv2d & In: 32, Out: 64, Kernel: \(3 \times 3\) & \(64 \times H/2 \times W/2\) \\
                           & ReLU   & - & \(64 \times 14 \times 14\) \\
                           & MaxPool2d & Kernel: \(2 \times 2\), Stride: 2 & \(64 \times H/4 \times W/4\) \\
                           & Flatten & - & \(1 \times (64 \times H/4 \times W/4) \) \\
                           & Linear  & In: \((64 \times H/4 \times W/4) \), Out: 128 & \(1 \times 128\) \\
                           & ReLU    & - & \(1 \times 128\) \\
                           & Linear  & In: 128, Out: 10 & \(1 \times 10\) \\
\bottomrule
\end{tabular}
}
\end{table}

\section{Discussions}
\label{sec:discussions}
\subsection{Efficiency}
The BSFL architecture enhances efficiency by selecting only the top-performing models for aggregation, resulting in a more optimized global model compared to SSFL. Additionally, the dynamic rotation of committee members ensures that the global model is trained on all local datasets in the system over time, effectively implementing a form of distributed k-fold cross-validation.

\subsection{Security and Stability}
Replacing the centralized FL server with a blockchain network enhances the system's reliability by eliminating the central FL server as a single point of failure. In this blockchain-based setup, all operations are carried out using smart contracts and collectively verified by a committee. This decentralized validation ensures secure and uninterrupted training processes, mitigating risks commonly associated with centralized FL servers. 

\subsection{Non-IID Data}

A common challenge in distributed learning is handling Non-Independent and Identically Distributed (Non-IID) local datasets. In simpler terms, each device might have biased or unbalanced data, for example, one device might predominantly store cat images, while another contains mostly dog images, resulting in local datasets that do not represent the overall data distribution.

To counter this challenge, our framework uses data from committee members for validation, exposing the global model to a broader, more varied set of data during evaluation. This cross-validation reduces the risk of overfitting by ensuring that the model is not overly specialized to the narrow distributions of individual nodes. By validating updates across diverse committee datasets, the system creates a more robust and generalized global model capable of handling diverse non-IID data effectively.
\begin{figure}[h]
  \centering
  \includegraphics[width=\linewidth]{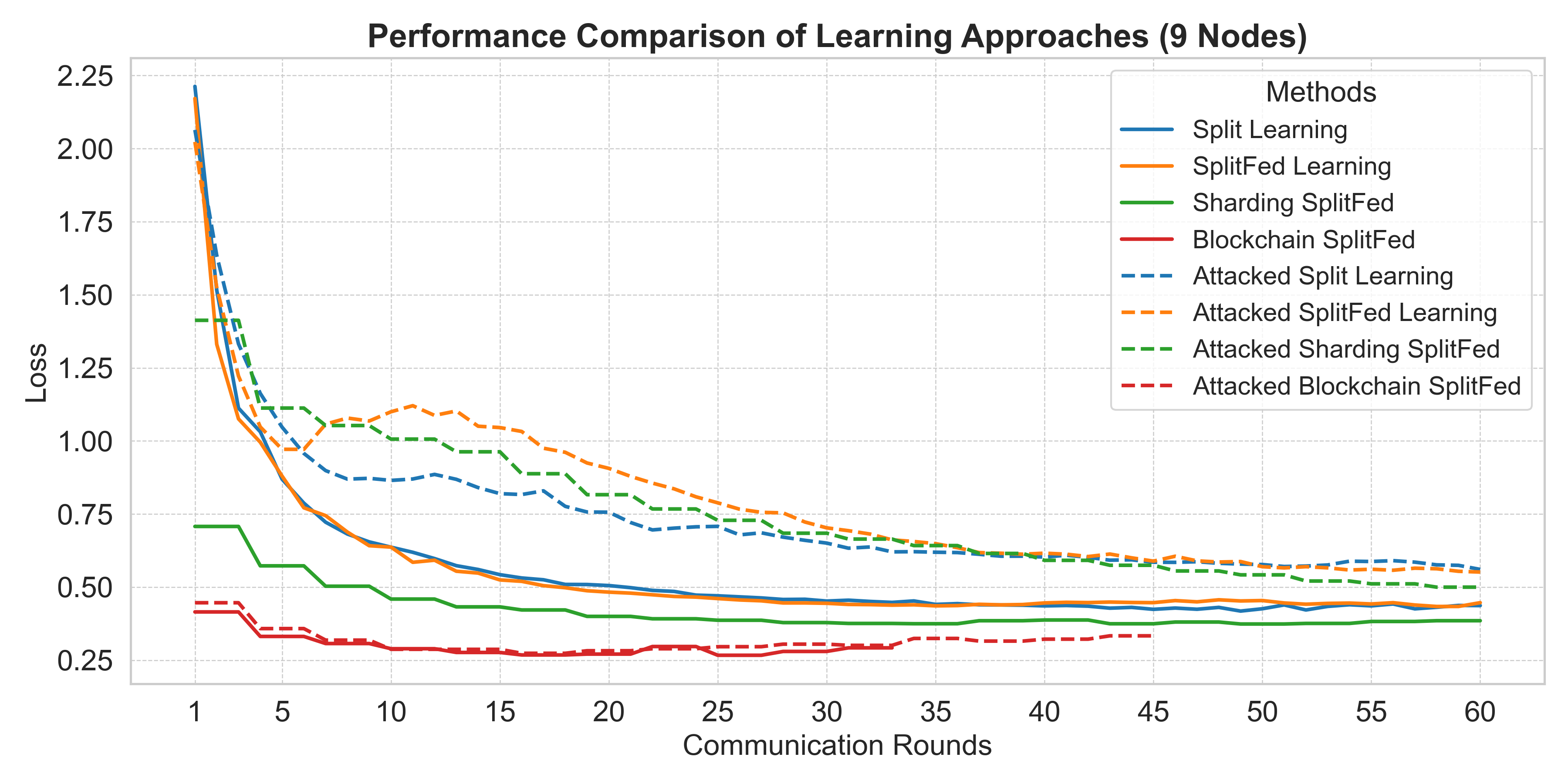}
  \caption{Performance comparison for 9 nodes in 60 training rounds.}
  \label{fig:all_9}
\end{figure}

\begin{figure}[h]
  \centering
  \includegraphics[width=\linewidth]{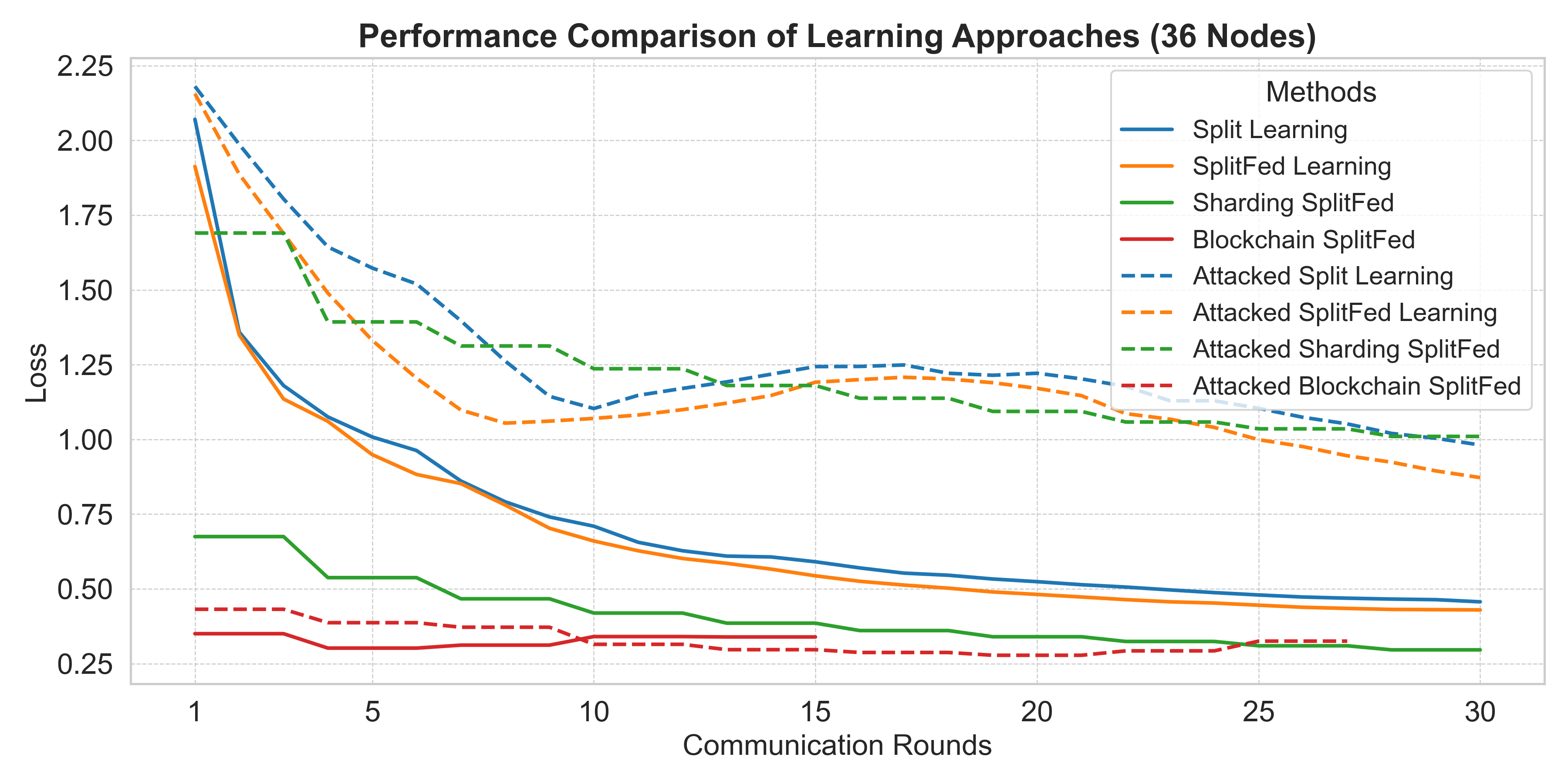}
  \caption{Performance comparison for 36 nodes in 30 training rounds.}
  \label{fig:all_36}
\end{figure}
\subsection{Committee Election}
In BSFL, we propose selecting committee members based on clients' scores from the previous round. Alternatively, a random selection approach could be employed to enhance model generalization by exposing the global model to a wider variety of datasets across multiple cycles.

Our current approach optimizes performance, stability, and fairness by selecting committee members from all clients in the previous round. However, this approach may appear to conflict with split learning’s goal of reducing the computational burden on resource-constrained clients. To resolve this, a more advanced committee selection algorithm could account for nodes’ computational capabilities. Restricting committee membership to nodes capable of performing shard server duties ensures that only suitable nodes are chosen. This refinement upholds the integrity of split learning while ensuring fairness and efficiency.

\begin{table*}[h]
    \centering
    \resizebox{\textwidth}{!}{%
        \begin{tabular}{l|r|r|r}
            \toprule
            \textbf{Approaches} & \textbf{Normal Test Loss} & \textbf{Attacked Test Loss} & \textbf{Avg. Round Time (min)} \\
            \midrule
            Split Learning (SL) & 0.456 & 0.981 & 37.6 \\
            SplitFed Learning (SFL) & 0.430 & 0.872 & 37.2 \\ 
            Sharding SplitFed Learning (SSFL) & 0.296 & 1.010 & 5.5 \\
            Blockchain-enabled SplitFed Learning (BSFL) & 0.339 & 0.325 & 33.7 \\
            \bottomrule
        \end{tabular}%
    }
    \vspace{0.5em}
    \caption{Performance comparison of learning approaches: normal and attacked test losses and average training times per round (36 nodes)}
    \label{tab:performance_comparison}
\end{table*}

\subsection{Malicious Nodes}

In SSFL, malicious nodes undermine system performance depending on their roles. When acting as clients, they may submit harmful model updates to degrade the global model’s accuracy. As FL servers, they can introduce noise into the global model or selectively favour specific clients, reducing model generalization. To mitigate these threats, we propose integrating a blockchain network with a committee consensus mechanism, eliminating the central FL server while implementing an evaluation framework for model updates. Below, we analyze the resilience of BSFL against malicious nodes in two key scenarios:

\begin{enumerate}
    \item Malicious Clients: Malicious nodes may submit ``poisonous" updates to degrade global model performance. Our proposed evaluation and committee selection algorithm mitigates such attacks by discarding harmful submissions and retaining only the top \( K \) most trustworthy model updates. As long as there are at least \( K \times C \) honest clients in the system (where \( C \) represents the number of clients per shard), the framework can maintain effective training and resist significant disruption from data-poisoning attacks.

\item Malicious Committee Members: Malicious nodes may infiltrate the committee and attempt to undermine the evaluation process. In a committee of \( N \) nodes, each model update receives a score from the remaining \( N - 1 \) committee members, with the final score calculated as the median via a smart contract. Malicious members may attempt to skew results by favoring inferior updates, but their impact is negligible unless they form a majority. To ensure resilience, BSFL framework requires at least \( \lfloor \frac{N}{2} \rfloor + 1 \) committee members to uphold evaluation integrity. Furthermore, to prevent the aggregation of harmful model updates, the number of selected models, \( K \), must remain less than \( \frac{N}{2} \).
\end{enumerate}
In summary, ensuring robust security in BSFL requires \( \lfloor \frac{N}{2} \rfloor + 1 \) honest committee members, at least \( K \times C \) honest clients, and selecting \( K \) values such that \( 2 < K < \frac{N}{2} \). However, the security requirements of BSFL are adaptable: in scenarios with minimal malicious activity, the value of \( K \) can be adjusted to prioritize efficiency over stringent security constraints.

\section{Experiments}
\label{sec:experiments}
\subsection{Settings and Normal Training}

To evaluate the performance of SSFL and BSFL, we implemented both frameworks using the Fashion MNIST dataset, which consists of 60,000 images across 10 distinct classes\footnote{The following link provides the implementation code in an anonymous GitHub repository to maintain the double-blinding constraint: \textbf{https://github.com/icdcs249/SSFL-BSFL}}. We define a node as any entity, client or SFL server, participating in a distributed learning procedure. This distinction is necessary since the clients and SFL servers are not static in frameworks such as BSFL. In our work, the experiments are conducted in two different settings: one with 9 nodes and the other with 36 nodes, to assess the effectiveness of our proposed solutions in both small-scale and large-scale scenarios. The local datasets for each node contain an equal number of images (6,666), but they are non-IID.

\begin{figure}[h]
  \centering
  \includegraphics[width=\linewidth]{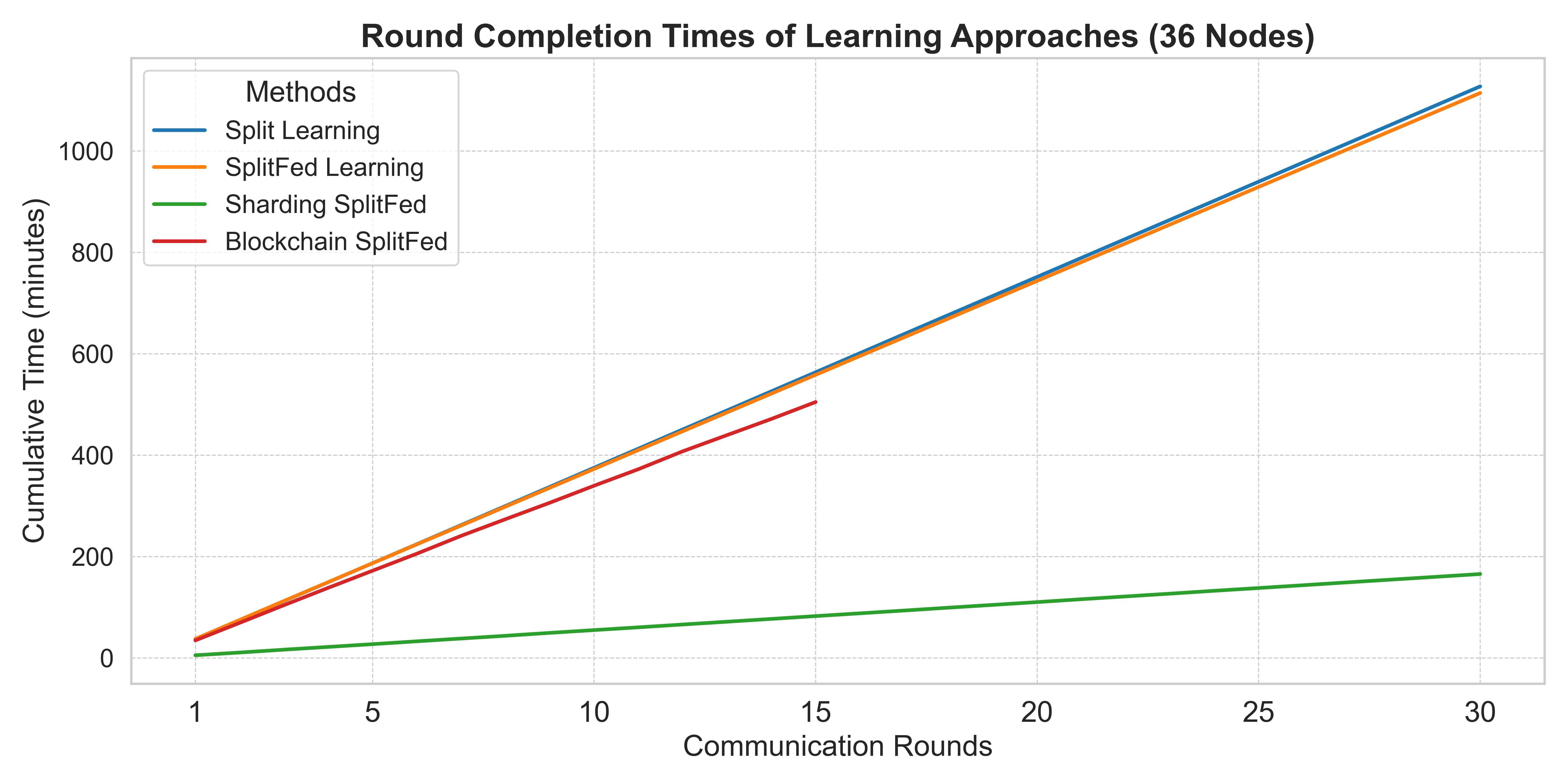}
  \caption{Round completion times for 36 nodes.}
  \label{fig:transmission}
\end{figure}

Additionally, we utilized the Hyperledger Fabric blockchain framework to implement the smart contracts (chaincodes) described in the BSFL architecture. The nodes in our framework are built by integrating PyTorch as the machine learning tool and Flask as the communication interface between the nodes and the blockchain network. Each node is implemented as a separate process in our experimental setup, which runs on an Intel Xeon(R) 4216 CPU with 16 cores and 32 threads (2 threads per core), along with an NVIDIA GeForce RTX 3080 GPU for accelerated training.

We compare the effectiveness of both SSFL and BSFL by conducting experiments under the same settings with SFL and SL. In each experiment, we train the global models, as described in Table \ref{table:model}, to classify Fashion MNIST images. In addition, the hyperparameters for these settings are fixed across all setups to ensure a precise evaluation of all approaches.

In the SL and SFL setups, one of the nodes (1 out of 9 or 1 out of 36) serves as the central server to oversee the training process. In SSFL and BSFL, we introduce three shards, each with two clients, for experiments involving nine nodes. When the number of nodes increases to 36, we adjust the configuration to six shards, each containing five clients. The value of $K$, which determines the number of top-performing shard SFL servers participating in the global model aggregation, is set to two and three for the 9-node and 36-node scenarios, respectively.

To prevent overfitting, we employ early stopping across all approaches. This is straightforward to implement in SSFL, SFL, and SL, as they involve a central node that supervises the training process. For BSFL, early stopping is achieved through the committee consensus mechanism, which halts training when the validation loss begins to deteriorate. 

Figures \ref{fig:all_9} and \ref{fig:all_36} show the validation loss of SL, SFL, SSFL, and BSFL across all training rounds. As evident in both figures, SSFL and BSFL outperform the earlier approaches, primarily due to the sharding mechanism employed in these frameworks. In both frameworks, the sharded architecture eradicates the imbalanced learning rate of split learning by aggregating the SFL server models, enhancing global model performance. Furthermore, BSFL demonstrates superior performance compared to SSFL by utilizing its committee consensus evaluation, which ensures that only the top-performing model updates contribute to the global model. Notably, BSFL requires fewer training rounds, as its enhanced performance leads the global model to reach optimal accuracy and begin overfitting the training data more quickly. Finally, Table \ref{tab:performance_comparison} compares the final test loss of all distributed learning algorithms. As shown, SSFL substantially enhances scalability and performance over SL and SFL due to its lower round time and test loss. Nevertheless, the test loss in the attacked settings proves SSFL's inability to tolerate data-poisoning threats.

\subsection{Under Malicious Attacks}
In our experiments, we assume that malicious nodes carry out data poisoning attacks by sending harmful updates to the central FL server, thereby undermining the performance of the global model. We conduct these attacks across all approaches to validate the effectiveness of our proposed solution. To evaluate the impact of data poisoning attacks, we test with varying proportions of malicious nodes, specifically $33\%$ and $47\%$ for the 9-node and 36-node setups, respectively. These varied attacker proportions allow us to evaluate the resilience and performance of our proposed frameworks under different threat levels. Notably, the 36-node experiment with $47\%$ attackers represents a scenario where the system is pushed to its limits, simulating the maximum number of attackers possible without breaching the $51\%$ threshold typically required for a successful blockchain takeover.

In BSFL, we further simulate a voting attack to assess the resilience of the committee consensus mechanism. In this scenario, malicious nodes, when selected as committee members, deliberately vote for the worst-performing model updates to disrupt the aggregation process. By examining the system’s ability to counter this disruption, we evaluate whether BSFL can still effectively aggregate the most beneficial model updates.

Figures \ref{fig:all_9} and \ref{fig:all_36} illustrate the performance of each algorithm under identical data-poisoning attacks. As shown in these figures, all algorithms except BSFL are adversely impacted by these attacks, lacking effective mechanisms to counteract them. In SL, SFL, and SSFL, data-poisoning attacks cause a substantial increase in validation loss and decline in the global model’s performance. BSFL, however, remains entirely unaffected due to its robust committee consensus mechanism, which effectively filters out malicious updates from adversarial nodes. The effectiveness of BSFL in mitigating data-poisoning attacks is also demonstrated by the test loss values presented in Table \ref{tab:performance_comparison}.

\subsection{Round Completion Time}

SSFL reduces transmission costs and communication overhead in SFL training by employing parallel shards, which distribute the workload across multiple nodes. BSFL incorporates blockchain technology to enhance the security of the system, which, in turn, increases the communication load due to committee member coordination and propagation of updates to and from the blockchain. In our experiments, we evaluate the round completion time of all training frameworks by measuring the time between the start and the end of each round and cycle, calculating the computation as well as the communication overhead for each round. This evaluation helps with identifying the most suitable distributed learning method by considering specific constraints, such as time, infrastructure, and other critical factors.

Figure \ref{fig:transmission} presents the transmission costs for each algorithm discussed. SSFL achieves significantly lower computation time compared to SL and SFL. BSFL incurs higher round completion time than SSFL because of its committee consensus mechanism and blockchain-based communication. However, BSFL compensates for this overhead with faster convergence, resulting in substantially lower overall training time compared to SL and SFL. Table \ref{tab:performance_comparison} displays the average round completion time for each algorithm, demonstrating that even without applying early stopping, BSFL maintains $11\%$ and $10\%$ lower average round completion time than SL and SFL, respectively, showcasing its efficiency and adaptability in secure distributed learning setups.

\section{Future Works}
In the following, we discuss three directions to enhance the scalability and evaluation mechanisms of the proposed solution by exploring other variations of split learning and adapting evaluation metrics.

\label{sec:future-works}
\subsection{SL variations}
An alternative approach to implementing SL and its variations, including SFL, involves splitting the model into three or more parts. In this configuration, the last few layers are also trained locally by clients, eliminating the need to share batch targets with the SL server. In this study, we only split the model into two parts; however, both SSFL and BSFL can be extended to support multi-part model splits. Such an extension could improve client-side privacy and reduce data exposure while maintaining the framework's scalability and performance. Nevertheless, this setup increases the computational power demanded from the clients since they must contribute to the training of more layers. This trade-off encourages further assessment from the distributed learning research community.

\subsection{Evaluation Metric}
Lastly, the evaluation mechanism can be adapted to metrics beyond traditional losses, especially for generative applications where labels are not required for assessing model performance. In these scenarios, metrics such as Feature Likelihood Divergence (FLD) \cite{jiralerspong2024feature} or Fréchet Inception Distance (FID) \cite{heusel2017gans} can be employed. These metrics are particularly relevant for evaluating nodes in tasks involving generative models, offering better insights into their performance.

\section{Conclusion}
\label{sec:conclusion}
In this paper, we introduced two new enhanced SplitFed Learning (SFL) frameworks called Sharded SplitFed Learning (SSFL) and Blockchian-enabled SplitFed Learning. SSFL addresses the scalability problem of SFL by distributing the workload of the SL server across multiple shards. Meanwhile, BSFL enhances the security and stability of the system by replacing the centralized server with a blockchain network. This network leverages a committee consensus mechanism with an evaluation process that prevents the aggregation of harmful model updates. We validated the effectiveness of our proposed frameworks through extensive experiments, comparing them with existing SL algorithms under various settings. Our results showcase the improvement in scalability, security, and convergence time for both SSFL and BSFL.
% Sara first review done!
\bibliography{ref}

\begin{thebibliography}{10}
\providecommand{\url}[1]{#1}
\csname url@samestyle\endcsname
\providecommand{\newblock}{\relax}
\providecommand{\bibinfo}[2]{#2}
\providecommand{\BIBentrySTDinterwordspacing}{\spaceskip=0pt\relax}
\providecommand{\BIBentryALTinterwordstretchfactor}{4}
\providecommand{\BIBentryALTinterwordspacing}{\spaceskip=\fontdimen2\font plus
\BIBentryALTinterwordstretchfactor\fontdimen3\font minus \fontdimen4\font\relax}
\providecommand{\BIBforeignlanguage}[2]{{%
\expandafter\ifx\csname l@#1\endcsname\relax
\typeout{** WARNING: IEEEtran.bst: No hyphenation pattern has been}%
\typeout{** loaded for the language `#1'. Using the pattern for}%
\typeout{** the default language instead.}%
\else
\language=\csname l@#1\endcsname
\fi
#2}}
\providecommand{\BIBdecl}{\relax}
\BIBdecl

\bibitem{duan2022combined}
Q.~Duan, S.~Hu, R.~Deng, and Z.~Lu, ``Combined federated and split learning in edge computing for ubiquitous intelligence in internet of things: State-of-the-art and future directions,'' \emph{Sensors}, vol.~22, no.~16, p. 5983, 2022.

\bibitem{mcmahan2017communication}
B.~McMahan, E.~Moore, D.~Ramage, S.~Hampson, and B.~A. y~Arcas, ``Communication-efficient learning of deep networks from decentralized data,'' in \emph{Artificial intelligence and statistics}.\hskip 1em plus 0.5em minus 0.4em\relax PMLR, 2017, pp. 1273--1282.

\bibitem{gupta2018distributed}
O.~Gupta and R.~Raskar, ``Distributed learning of deep neural network over multiple agents,'' \emph{Journal of Network and Computer Applications}, vol. 116, pp. 1--8, 2018.

\bibitem{vepakomma2018split}
P.~Vepakomma, O.~Gupta, T.~Swedish, and R.~Raskar, ``Split learning for health: Distributed deep learning without sharing raw patient data,'' \emph{arXiv preprint arXiv:1812.00564}, 2018.

\bibitem{ZHANG2021106775}
\BIBentryALTinterwordspacing
C.~Zhang, Y.~Xie, H.~Bai, B.~Yu, W.~Li, and Y.~Gao, ``A survey on federated learning,'' \emph{Knowledge-Based Systems}, vol. 216, p. 106775, 2021. [Online]. Available: \url{https://www.sciencedirect.com/science/article/pii/S0950705121000381}
\BIBentrySTDinterwordspacing

\bibitem{Batool2022BlockFeST}
Z.~Batool, K.~Zhang, Z.~Zhu, S.~Aravamuthan, and U.~Aivodji, ``Block-fest: A blockchain-based federated anomaly detection framework with computation offloading using transformers,'' in \emph{2022 IEEE 1st Global Emerging Technology Blockchain Forum: Blockchain \& Beyond (iGETblockchain)}, 2022, pp. 1--6.

\bibitem{Gao2020endtoend}
Y.~Gao, M.~Kim, S.~Abuadbba, Y.~Kim, C.~Thapa, K.~Kim, S.~A. Camtep, H.~Kim, and S.~Nepal, ``End-to-end evaluation of federated learning and split learning for internet of things,'' in \emph{2020 International Symposium on Reliable Distributed Systems (SRDS)}, 2020, pp. 91--100.

\bibitem{ceballos2020splitnn}
I.~Ceballos, V.~Sharma, E.~Mugica, A.~Singh, A.~Roman, P.~Vepakomma, and R.~Raskar, ``Splitnn-driven vertical partitioning,'' \emph{arXiv preprint arXiv:2008.04137}, 2020.

\bibitem{thapa2022splitfed}
C.~Thapa, P.~C.~M. Arachchige, S.~Camtepe, and L.~Sun, ``Splitfed: When federated learning meets split learning,'' in \emph{Proceedings of the AAAI Conference on Artificial Intelligence}, vol.~36, no.~8, 2022, pp. 8485--8493.

\bibitem{oh2022locfedmix}
S.~Oh, J.~Park, P.~Vepakomma, S.~Baek, R.~Raskar, M.~Bennis, and S.-L. Kim, ``Locfedmix-sl: Localize, federate, and mix for improved scalability, convergence, and latency in split learning,'' in \emph{Proceedings of the ACM Web Conference 2022}, 2022, pp. 3347--3357.

\bibitem{pal2021server}
S.~Pal, M.~Uniyal, J.~Park, P.~Vepakomma, R.~Raskar, M.~Bennis, M.~Jeon, and J.~Choi, ``Server-side local gradient averaging and learning rate acceleration for scalable split learning,'' \emph{arXiv preprint arXiv:2112.05929}, 2021.

\bibitem{oh2023mix2sfl}
S.~Oh, H.~Nam, J.~Park, P.~Vepakomma, R.~Raskar, M.~Bennis, and S.-L. Kim, ``Mix2sfl: Two-way mixup for scalable, accurate, and communication-efficient split federated learning,'' \emph{IEEE Transactions on Big Data}, 2023.

\bibitem{li2021bflc}
Y.~Li, C.~Chen, N.~Liu, H.~Huang, Z.~Zheng, and Q.~Yan, ``A blockchain-based decentralized federated learning framework with committee consensus,'' \emph{IEEE Network}, vol.~35, no.~1, pp. 234--241, 2021.

\bibitem{qu2022blockchain}
Y.~Qu, M.~P. Uddin, C.~Gan, Y.~Xiang, L.~Gao, and J.~Yearwood, ``Blockchain-enabled federated learning: A survey,'' \emph{ACM Computing Surveys}, vol.~55, no.~4, pp. 1--35, 2022.

\bibitem{pasquini2021unleashing}
D.~Pasquini, G.~Ateniese, and M.~Bernaschi, ``Unleashing the tiger: Inference attacks on split learning,'' in \emph{Proceedings of the 2021 ACM SIGSAC Conference on Computer and Communications Security}, 2021, pp. 2113--2129.

\bibitem{kairouz2021advances}
P.~Kairouz, H.~B. McMahan, B.~Avent, A.~Bellet, M.~Bennis, A.~N. Bhagoji, K.~Bonawitz, Z.~Charles, G.~Cormode, R.~Cummings \emph{et~al.}, ``Advances and open problems in federated learning,'' \emph{Foundations and trends{\textregistered} in machine learning}, vol.~14, no. 1--2, pp. 1--210, 2021.

\bibitem{lyu2020threats}
L.~Lyu, H.~Yu, and Q.~Yang, ``Threats to federated learning: A survey,'' \emph{arXiv preprint arXiv:2003.02133}, 2020.

\bibitem{shokri2017membership}
R.~Shokri, M.~Stronati, C.~Song, and V.~Shmatikov, ``Membership inference attacks against machine learning models,'' in \emph{2017 IEEE symposium on security and privacy (SP)}.\hskip 1em plus 0.5em minus 0.4em\relax IEEE, 2017, pp. 3--18.

\bibitem{dwork2006differential}
C.~Dwork, ``Differential privacy,'' in \emph{International colloquium on automata, languages, and programming}.\hskip 1em plus 0.5em minus 0.4em\relax Springer, 2006, pp. 1--12.

\bibitem{li2020fl}
T.~Li, A.~K. Sahu, A.~Talwalkar, and V.~Smith, ``Federated learning: Challenges, methods, and future directions,'' \emph{IEEE Signal Processing Magazine}, vol.~37, no.~3, pp. 50--60, 2020.

\bibitem{kim2020blockfl}
H.~Kim, J.~Park, M.~Bennis, and S.-L. Kim, ``Blockchained on-device federated learning,'' \emph{IEEE Communications Letters}, vol.~24, no.~6, pp. 1279--1283, 2020.

\bibitem{Sokhankhosh2024PoCL}
A.~Sokhankhosh and S.~Rouhani, ``Proof-of-collaborative-learning: A multi-winner federated learning consensus algorithm,'' in \emph{2024 IEEE International Conference on Blockchain (Blockchain)}, 2024, pp. 370--377.

\bibitem{truex2019hybrid}
S.~Truex, N.~Baracaldo, A.~Anwar, T.~Steinke, H.~Ludwig, R.~Zhang, and Y.~Zhou, ``A hybrid approach to privacy-preserving federated learning,'' in \emph{Proceedings of the 12th ACM workshop on artificial intelligence and security}, 2019, pp. 1--11.

\bibitem{qu2021proof}
X.~Qu, S.~Wang, Q.~Hu, and X.~Cheng, ``Proof of federated learning: A novel energy-recycling consensus algorithm,'' \emph{IEEE Transactions on Parallel and Distributed Systems}, vol.~32, no.~8, pp. 2074--2085, 2021.

\bibitem{Lu2020PoTQ}
Y.~Lu, X.~Huang, Y.~Dai, S.~Maharjan, and Y.~Zhang, ``Blockchain and federated learning for privacy-preserved data sharing in industrial iot,'' \emph{IEEE Transactions on Industrial Informatics}, vol.~16, no.~6, pp. 4177--4186, 2020.

\bibitem{Yuan2021ChainsFL}
S.~Yuan, B.~Cao, M.~Peng, and Y.~Sun, ``Chainsfl: Blockchain-driven federated learning from design to realization,'' in \emph{2021 IEEE Wireless Communications and Networking Conference (WCNC)}, 2021, pp. 1--6.

\bibitem{Alief2023FLB2}
R.~N. Alief, M.~A. Paramartha~Putra, A.~Gohil, J.-M. Lee, and D.-S. Kim, ``Flb2: Layer 2 blockchain implementation scheme on federated learning technique,'' in \emph{2023 International Conference on Artificial Intelligence in Information and Communication (ICAIIC)}, 2023, pp. 846--850.

\bibitem{zhang2023blockchain}
Y.~Zhang, Y.~Tang, Z.~Zhang, M.~Li, Z.~Li, S.~Khan, H.~Chen, and G.~Cheng, ``Blockchain-based practical and privacy-preserving federated learning with verifiable fairness,'' \emph{Mathematics}, vol.~11, no.~5, p. 1091, 2023.

\bibitem{Sai2024Ablockchain}
S.~Sai and V.~Chamola, ``A blockchain-enabled split learning framework with a novel client selection method for collaborative learning in smart healthcare,'' \emph{IEEE Transactions on Consumer Electronics}, pp. 1--1, 2024.

\bibitem{jiralerspong2024feature}
M.~Jiralerspong, J.~Bose, I.~Gemp, C.~Qin, Y.~Bachrach, and G.~Gidel, ``Feature likelihood score: Evaluating the generalization of generative models using samples,'' \emph{Advances in Neural Information Processing Systems}, vol.~36, 2024.

\bibitem{heusel2017gans}
M.~Heusel, H.~Ramsauer, T.~Unterthiner, B.~Nessler, and S.~Hochreiter, ``Gans trained by a two time-scale update rule converge to a local nash equilibrium,'' \emph{Advances in neural information processing systems}, vol.~30, 2017.

\end{thebibliography}
\bibliographystyle{IEEEtran}

\end{document}